\documentclass[prd,eqsecnum,twocolumn,amsfonts,showpacs]{revtex4}

\usepackage{graphicx}

\usepackage{bm}

\def\fsl#1{\setbox0=\hbox{$#1$}           
   \dimen0=\wd0                                 
   \setbox1=\hbox{/} \dimen1=\wd1               
   \ifdim\dimen0>\dimen1                        
      \rlap{\hbox to \dimen0{\hfil/\hfil}}      
      #1                                        
   \else                                        
      \rlap{\hbox to \dimen1{\hfil$#1$\hfil}}   
      /                                         
   \fi}                                         %

\newcommand{\beq}{\begin{equation}}
\newcommand{\eeq}{\end{equation}}
\newcommand{\beqs}{\begin{eqnarray}}
\newcommand{\eeqs}{\end{eqnarray}}
\newcommand{\lsim}{\mathrel{\raisebox{-
.6ex}{$\stackrel{\textstyle<}{\sim}$}}}

\begin{document}

\title{Study of the Change from Walking to Non-Walking Behavior \\
in a Vectorial Gauge Theory as a Function of $N_f$} 

\author{Masafumi Kurachi}

\author{Robert Shrock}

\affiliation{
C.N. Yang Institute for Theoretical Physics \\
State University of New York \\
Stony Brook, NY 11794}

\begin{abstract}

We study a vectorial gauge theory with gauge group ${\rm SU}(N_c)$ and a
variable number, $N_f$, of massless fermions in the fundamental representation
of this group.  Using approximate solutions of Schwinger-Dyson and
Bethe-Salpeter equations, we calculate meson masses and investigate how these
depend on $N_f$.  We focus on the range of $N_f$ extending from near the
boundary with a non-Abelian Coulomb phase, where the theory exhibits a slowly
running (``walking'') gauge coupling, toward smaller values where the theory
has non-walking behavior. Our results include determinations of the masses of
the lowest-lying flavor-adjoint mesons with $J^{PC}=0^{-+}$, $1^{--}$,
$0^{++}$, and $1^{++}$ (the generalized $\pi$, $\rho$, $a_0$, and $a_1$).
Related results are given for flavor-singlet mesons and for the generalization
of $f_\pi$.  These results give insight into the change from walking to
non-walking behavior in a gauge theory, as a function of $N_f$. 

\end{abstract}



\maketitle

\vspace{16mm}

\newpage
\pagestyle{plain}
\pagenumbering{arabic}

\section{Introduction}

We consider a $(3+1)$-dimensional vectorial gauge theory (at zero temperature
and chemical potential) with the gauge group SU($N_c$) and $N_f$ massless
fermions transforming according to the fundamental representation of this
group.  For $N_c=3$, if one took $N_f=2$, this would be an approximation to
actual quantum chromodynamics (QCD) with just the $u$ and $d$ quarks, since
their current quark masses are small compared with the scale $\Lambda_{QCD}
\simeq 400$ MeV.  We restrict here to the range $N_f < (11/2)N_c$ for which the
theory is asymptotically free.  An analysis using the two-loop beta function
and Schwinger-Dyson equation (reviewed below) leads to the inference that for
$N_f$ in this range, the theory includes two phases: (i) for $0 \le N_f \le
N_{f,cr}$ a phase with confinement and spontaneous chiral symmetry breaking
(S$\chi$SB); and (ii) for $N_{f,cr} \le N_f \le (11/2)N_c$ a non-Abelian
Coulomb phase with no confinement or spontaneous chiral symmetry breaking. We
shall refer to $N_{f,cr}$, the critical value of $N_f$, as the boundary of the
non-Abelian Coulomb (conformal) phase \cite{bz}.  Here we take electroweak
interactions to be turned off.  We denote the fermions as $f_i^a$ with
$a=1,...,N_c$ and $i=1,...,N_f$.  The theory has an ${\rm SU}(N_f)_L \times
{\rm SU}(N_f)_R \times {\rm U}(1)_V$ global symmetry (the U(1)$_A$ being
explicitly broken by instantons), which is spontaneously broken to ${\rm
SU}(N_f)_V \times {\rm U}(1)_V$ by the formation of a bilinear fermion
condensate.

For $N_f$ slightly less than $N_{f,cr}$, the theory exhibits an approximate
infrared (IR) fixed point with resultant walking behavior.  That is, as the
energy scale $\mu$ decreases from large values, $\alpha=g^2/(4\pi)$ ($g$ being
the SU($N_c$) gauge coupling) grows to be O(1) at a scale $\Lambda$, but
increases only rather slowly as $\mu$ decreases below this scale, so that there
is an extended interval in energy below $\Lambda$ where $\alpha$ is large, but
slowly varying.  Associated with this slowly running behavior, the resultant
dynamically generated fermion mass, $\Sigma$, is much smaller than $\Lambda$.
In addition to its intrinsic field-theoretic interest, this walking behavior
has played an important role in theories of dynamical electroweak symmetry
breaking \cite{wtc1}-\cite{chipt3}.  As $N_f$ approaches $N_{f,cr}$ from below,
quantities with dimensions of mass vanish continuously; i.e., the chiral phase
transition separating phases (i) and (ii) is continous.  Recently, meson masses
and other quantities such as the generalized pseudoscalar decay constant
$f_\pi$ were calculated in the walking limit of an SU($N_c$) gauge theory
\cite{mm}.

In the present paper we shall investigate how meson masses and other quantities
change as one decreases $N_f$ below $N_{f,cr}$, moving away from the boundary,
as a function of $N_f$, between phases (i) and (ii), deeper into the confined
phase.  Our paper is thus a study of the change (crossover) between the walking
behavior that occurs near to this boundary, and the non-walking behavior that
occurs for smaller $N_f$.  In a non-walking (asymptotically free, confining)
theory such as real QCD, as the energy scale $\mu$ decreases through $\Lambda$,
$\alpha$ increases rapidly through values of order unity, triggering
spontaneous chiral symmetry breaking on this scale, so that $\Sigma \sim
\Lambda$.  This is quite different from a theory with walking, in which $\Sigma
\ll \Lambda$.  Our basic calculational methods are essentially the same as
those employed in Ref. \cite{mm}, i.e., we use the Schwinger-Dyson (SD)
equation to compute the dynamical fermion mass $\Sigma$ (generalized
constituent quark mass) and then insert this into the Bethe-Salpeter (BS)
equation to obtain the masses of the low-lying mesons.  We restrict to an
interval of $N_f$ values for which the theory has an infrared fixed point (as
calculated from the beta function, to be discussed further below).  The reason
for this is that it makes our calculations more robust since for our interval
of $N_f$ we can avoid having to introduce a cutoff on the growth of $\alpha$ in
the infrared.  If one uses Schwinger-Dyson and Bethe-Salpeter equations to
explore a region of $N_f$ where the beta function does not have an infrared
fixed point, one must use such an IR cutoff, which leads to cutoff-dependence
of the results.  For definiteness, we shall take $N_c=3$; however, as will be
seen, $N_c$ only enters indirectly, via the dependence of the value of the
infrared fixed point $\alpha_*$ (eq. (\ref{alfcrit}) below) on $N_c$.  Hence,
our findings may also be applied in a straightforward way, with appropriate
changes in the value of $\alpha_*$, to an SU($N_c$) gauge theory with a
different value of $N_c$.

We mention some background and related work.  Many studies have investigated
the hadron mass spectrum for QCD with $N_f=2$ light quarks.  Among the earliest
were static quark models \cite{qm}, and bag models\cite{mitbag,mitbag_pwave}.
Lattice gauge theory has provided an especially powerful method \cite{lat}. The
Schwinger-Dyson and Bethe-Salpeter equations have been used for many years to
study spontaneous chiral symmetry breaking and relativistic bound states in
field theories (a partial list of papers and reviews includes \cite{wtc2}-
\cite{mm}, \cite{bs}-\cite{HY}).  In particular, the
Bethe-Salpeter equation has been used to calculate meson masses in QCD
\cite{abkmn}-\cite{marisroberts03}.  For the walking limit, in addition to
Refs. \cite{mm}, these methods have also been used in connection with spectral
function sum rules to study the $\pi^+ - \pi^0$ mass difference \cite{pimdif}
and the $S$ parameter (equivalently, the chiral Lagrangian coefficient
$L_{10}$) \cite{HKY}.

This paper is organized as follows. In Section II we review some background
material concerning the beta function, approximate infrared fixed point, and
walking behavior.  Section III includes a discussion of the Schwinger-Dyson
equation and our solution of it, as well as our calculation of the pseudoscalar
decay constant $f_P$, the generalization of $f_\pi$.  In Section IV we present
our calculation of meson masses using the Bethe-Salpeter equation.  Section V
contains some further remarks and our conclusions.

\section{Preliminaries}

In order to study meson masses and other quantities as one moves away from the
boundary between the confined phase with spontaneous chiral symmetry breaking
and the non-Abelian Coulomb phase, it is first necessary to know as accurately
as possible where this boundary lies, as a function of $N_f$, i.e., to know the
value of $N_{f,cr}$.  We first review the estimate \cite{chipt3} based on using
the two-loop SU($N_c$) beta function \cite{b0,b1}
\beq
\beta = \frac{d \alpha}{dt} = - \frac{\alpha^2}{2\pi}\left ( b_0 +
\frac{b_1}{4\pi}\alpha + O(\alpha^2) \right ) \ ,
\label{beta}
\eeq
where $t=\ln \mu$, with $\mu$ the energy scale. The two terms listed are
scheme-independent.  (Two higher-order terms in $\beta$ have been calculated
but are scheme-dependent; inclusion of these does not significantly affect our
results.) For the relevant case of an asymptotically free theory, $b_0 > 0$ so
that an infrared fixed point exists if $b_1 < 0$.  This coefficient $b_1$ is
positive for $0 \le N_f \le N_{f,IR}$, where
\beq
N_{f,IR}=\frac{34N_c^3}{13N_c^2-3}
\label{nfir}
\eeq
and negative for larger $N_f$.  For $N_c=3$, $N_{f,IR} \simeq 8.1$
\cite{integer}.  The value of $\alpha$ at this IR fixed point, denoted
$\alpha_*$, is given by $\alpha_* = -4\pi b_0/b_1$.  Substituting the known
values of these terms, one has 
\beq
\alpha_* = \frac{-4\pi(11N_c -2N_f)}{34N_c^2-13N_cN_f+3N_c^{-1}N_f} \ . 
\label{alpha_irfp}
\eeq

Solving eq. (\ref{alpha_irfp}) for $N_f$ in terms of $\alpha_*$ yields 
\beq
N_f=\frac{2N_c^2[17N_c(\alpha_*/\pi)+22]}{(13N_c^2-3)(\alpha_*/\pi)+8N_c} \ . 
\label{nfsol}
\eeq
In the one-gluon exchange approximation, the Schwinger-Dyson gap equation for
the inverse propagator of a fermion transforming according to the
representation $R$ of SU($N_c$) has a nonzero solution for the dynamically
generated fermion mass, which is an order parameter for spontaneous chiral
symmetry breaking, if $\alpha \ge \alpha_{cr}$, where $\alpha_{cr}$ is given by
\beq
\frac{3 \alpha_{cr} C_2(R)}{\pi} = 1,
\label{alfcritcondition}
\eeq
and $C_2(R)$ denotes the quadratic Casimir invariant for the representation $R$
\cite{casimir}.  Using $C_2(fund.) \equiv C_{2f}=(N_c^2-1)/(2N_c)$ for the
fundamental representation yields
\beq
\alpha_{cr} = \frac{2\pi N_c}{3(N_c^2-1)} \ . 
\label{alfcrit}
\eeq
For the case $N_c=3$ that we use for definiteness here, eq. (\ref{alfcrit})
gives $\alpha_{cr} = \pi/4 \simeq 0.79$.  To estimate $N_{f,cr}$, one solves
the equation $\alpha_* = \alpha_{cr}$, yielding the result \cite{chipt3}
\beq
N_{f,cr} = \frac{2N_c(50N_c^2-33)}{5(5N_c^2-3)} \ . 
\label{nfcr}
\eeq
For $N_c=3$ this gives $N_{f,cr} \simeq 11.9$.  These estimates are only rough,
in view of the strongly coupled nature of the physics.  Effects of higher-order
gluon exchanges have been studied in Ref. \cite{alm}.  These calculations are
semi-perturbative and do not include instanton effects.  It is known that
instantons enhance the formation of the bilinear fermion condensates
\cite{instantons}, which suggests that their inclusion would expand the phase
with confinement and spontaneous chiral symmetry breaking, i.e., would increase
$N_{f,cr}$ somewhat relative to the value obtained from the two-loop beta
function and gap equation.  In principle, lattice gauge simulations provide a
way to determine $N_{f,cr}$, but the groups that have studied this have
not reached a consensus \cite{iwasaki}-\cite{mawhinney}.  

In our analysis, what we actually vary is the value of the approximate IR fixed
point $\alpha_*$, which depends parametrically on $N_f$.  Thus, although our SD
and BS equations are semi-perturbative, the analysis is self-consistent in the
sense that our $\alpha_{cr}$ really is the value at which, in our
approximation, one passes from the confinement phase to the non-Abelian Coulomb
phase, and our values of $\alpha$ do span the interval over which there is a
crossover from walking to QCD-like (i.e., non-walking) behavior. 

We elaborate here on the origin of the walking behavior.  Since the theory is
asymptotically freee, it follows that as the energy scale $\mu$ decreases from
values $\gg \Lambda$, $\alpha$ increases.  If $N_f < N_{f,IR}$, there is no
perturbative IR fixed point.  If $N_{f,IR} < N_f < N_{f,cr}$, as the energy
scale decreases toward zero, the coupling $\alpha$ approaches $\alpha_*$, which
is larger than $\alpha_{cr}$.  The coupling $\alpha_*$ is only an approximate
IR fixed point since, as $\alpha$ increases past $\alpha_{cr}$ and the fermion
condensate forms, the fermions gain a dynamical mass $\Sigma$ so that in the
low-energy effective theory applicable for energy scales $\mu < \Sigma$, one
integrates out these fermions and is left with a pure gluonic SU($N_c$) theory
with a different beta function, such that $\alpha$ increases further. If $N_f >
N_{f,cr}$ (and smaller than $(11/2)N_c$), the theory is in the non-Abelian
Coulomb phase and $\alpha_*$ is an exact IR fixed point.  In the case that
$N_f$ is only slightly less than $N_{f,cr}$, or equivalently, $\alpha_*$ is
only slightly greater than $\alpha_{cr}$, it follows that as the energy scale
decreases and $\alpha$ approaches $\alpha_*$ from below, the rate of increase
of $\alpha$, i.e., $|\beta|$, decreases, so that the theory has a large
coupling $\alpha \sim O(1)$ which, however, runs very slowly.

As is evident from the above results, decreasing $N_f$ below $N_{f,cr}$ has the
effect of increasing $\alpha_*$ and thus moving the theory deeper in the phase
with confinement and spontaneous chiral symmetry breaking, away from the
boundary with the non-Abelian Coulomb phase.  This is the key parametric
dependence that we shall use for our study. Our aim is to investigate how meson
masses and other observable quantities depend on $N_f$ in the crossover region;
operationally, what we actually vary is $\alpha_*$.  In Ref. \cite{mm} the
range of $\alpha_*$ used for the calculation of meson masses was chosen to be
$0.89 \le \alpha_* \le 1.0$, an interval where there is pronounced walking
behavior.  For the case $N_c=3$ considered in Ref. \cite{mm} and here, given
the above-mentioned value, $\alpha_{cr}=\pi/4$, it follows that this lower
limit, $\alpha_*=0.89$, is about 12 \% greater than this critical coupling.
The reason for this choice of lower limit on $\alpha_*$ was that the hadron
masses become exponentially small relative to the scale $\Lambda$ as
$\alpha_* - \alpha_{cr} \to 0^+$, rendering numerical evaluations of the
relevant integrals increasingly difficult in this limit.  For our study of the
shift away from walking behavior we consider an interval extending to larger
couplings, from $\alpha_*=1.0$ to $\alpha_*=2.5$.  Our upper limit is chosen in
order for the ladder approximation used in our solutions of the Schwinger-Dyson
and Bethe-Salpeter equations to have reasonable reliability.  From
eq. (\ref{nfsol}) it follows that $\alpha_*=0.89$ corresponds to $N_f=11.65$,
about 2 \% less than $N_{f,cr}$.  For a coupling as large as $\alpha_* = 2.5$,
the semi-perturbative methods used to derive eqs. (\ref{alpha_irfp}) and
(\ref{nfsol}) are subject to large corrections from higher-order perturbative,
and from nonperturbative, contributions; recognizing this, the above upper
limit of $\alpha_*$ corresponds formally to $N_f \simeq 9.8$, a roughly 20 \%
reduction from $N_{f,cr}=11.9$.

Since the chiral transition which occurs as $N_f$ increases through $N_{f,cr}$
is second-order (continuous), and since there is no spontaneous chiral symmetry
breaking in the non-Abelian Coulomb phase, it follows that as $N_f \nearrow
N_{f,cr}$, (i) the masses of all hadron states vanish continuously; and (ii)
hadron states that are related to each other by a parity reflection become
degenerate.

\section{Schwinger-Dyson Equation}

We first use the Schwinger-Dyson equation for the fermion propagator to
calculate the dynamically generated mass $\Sigma$ of this fermion.  This
extends the calculation of these quantities in Ref. \cite{mm} to smaller $N_f$
and, accordingly, larger $\alpha_*$.  The inverse fermion propagator is
$S_f(p)^{-1} = A(p^2) \fsl{p} - B(p^2)$.  We approximate the full
Schwinger-Dyson equation by using an effective running coupling and the
lowest-order gluon propagator:
\beq
S_f(p)^{-1} - \fsl{p} = -i C_{2f} \int \frac{d^4 q}{(2\pi)^4}
\bar{g}^2(p,q) \, D_{\mu\nu}(p-q) \,  \gamma^\mu \, S_f(q) \, \gamma^\nu
\label{sdeqgen}
\eeq
We use the Landau gauge for the gluon propagator $D_{\mu\nu}(k)$, i.e.,
$D_{\mu\nu} = (-g_{\mu\nu}+k_\mu k_\nu/k^2)/k^2$ because this simplifies the
calculation.  The physical results are, of course, gauge-invariant (e.g.,
\cite{alm}).  Equation (\ref{sdeqgen}) yields two separate equations for
$A(p^2)$ and $B(p^2)$.  As in Ref. \cite{mm}, we make the ansatz for the
running coupling, after Euclidean rotation,
\beq
\alpha(p_E,q_E) = \alpha(p_E^2+q_E^2) \ , 
\label{gsqform}
\eeq
where the subscript denotes Euclidean.  Since $\alpha$ would naturally depend
on the gluon momentum squared, $(p-q)^2 = p^2+q^2-2p \cdot q$, the
functional form (\ref{gsqform}) amounts to dropping the scalar product term,
$-2p \cdot q$.  This is a particularly reasonable approximation in the case of
a walking gauge theory because most of the contribution to the integral
(\ref{sdeqgen}) comes from a region of Euclidean momenta where $\alpha$ is
nearly constant.  Hence, the shift upward or downward due to the $-2p \cdot q$
term in the argument of $\alpha$ has very little effect on the value of this
coupling for the range of momenta that make the most important contribution to
the integral. The approximation (\ref{gsqform}) enables one to carry out the
angular integration, obtaining the results $A(p_E^2)=1$ and, for $B(p_E^2)
\equiv \Sigma(p_E^2)$, setting $x \equiv p_E^2$ and $y \equiv q_E^2$, 
\beq
\Sigma(x) = \frac{3 C_{2f}}{4 \pi} \int_0^\infty y \, dy
\frac{\alpha(x+y) \, \Sigma(y)}{\max(x,y) \, [y + \Sigma^2(y)]} \ . 
\label{sdeq}
\eeq
In terms of the momentum-scale-dependent fermion mass $\Sigma(p_E^2)$, we 
define the dynamical mass $\Sigma$ as
\beq
\Sigma \equiv \Sigma (p_E^2=\Sigma^2) \ . 
\label{sigdef}
\eeq

As noted above, in the walking region, over most of the range of integration
over $q_E$ in eq. (\ref{sdeq}) below $\Lambda$, the running coupling $\alpha$
is approximately constant and equal to its fixed-point value, $\alpha_*$ (see
Fig. 2 of Ref. \cite{mm}).  This means that in the walking region one does not
have to introduce any infrared cutoff on the growth of $\alpha$, as was
necessary in earlier studies of the Schwinger-Dyson and Bethe-Salpeter
equations for regular QCD \cite{higashijima}, \cite{miranskyrev}, \cite{abkm}.
In the interval $q_E \lsim \Sigma \ll \Lambda$, the fermions decouple, having
gained dynamical masses $\Sigma$, and in this low-energy theory with the
fermions integrated out, the resultant $\alpha$ would evolve away from
$\alpha_*$ as calculated via the perturbative beta function.  However, since
this dynamical mass scale is much smaller than $\Lambda$ in a walking theory,
it follows that this lowest range of the integration over $q_E$ makes a
negligibly small contribution to the entire integral.  One can thus employ the
approximation of using the same functional form for $\alpha$ down to $q_E=0$ in
the integral.  This convenient feature does not hold if $N_f$ decreases very
far below $N_{f,cr}$, i.e., $\alpha_*$ increases too far above $\alpha_{cr}$.

Having made the approximation of using the same functional form for $\alpha$
for $k_E$ in the range $0 \le k_E \le \Sigma$ as in the range $k_E >
\Sigma$, and solving for $\alpha(k_E)$ from the two-loop beta function, one
finds that, in terms of the variable $\ln(k_E/\Lambda)$, it increases rather
quickly from small values to values of O(1) as $k_E$ decreases below $\Lambda$.
This motivates an additional simplification, namely approximating $\alpha(k_E)$
as the step function,
\beq
\alpha(k_E) = \alpha_*\theta(\Lambda - k_E) \ . 
\label{stepfun}
\eeq

Then as $\alpha_* \searrow \alpha_{cr}$, if one also approximates the
denominator of the fermion propagator in eq. (\ref{sdeq}) as 
$(q_E^2+\Sigma^2)$, i.e., one sets $\Sigma(q_E^2) = \Sigma$ in this
denominator, then the solution is \cite{chipt2}-\cite{chipt3}
\beq
\Sigma = c \Lambda \, \exp
\bigg [ -\pi \Big ( \frac{\alpha_*}{\alpha_{cr}} - 1 \Big )^{-1/2} \bigg ] \ , 
\label{sigsol}
\eeq
where $c$ is a constant. 

\begin{figure}
  \begin{center}
    \includegraphics[height=6cm]{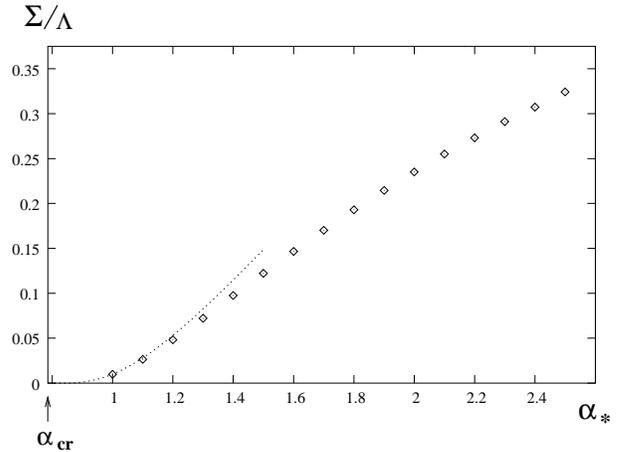}
  \end{center}
\caption{Numerical solutions for $\Sigma$, for several values of $\alpha_*$
(indicated by $\Diamond$). By way of comparison, we show, as the dotted curve,
the solution (\ref{sigsol}) with $c=4.0$ derived from a fit to the results in
the interval $0.89 \le \alpha_* \le 1.0$.  See text for further discussion.}
\label{Sigma_comp}
\end{figure}

One next discretizes the Schwinger-Dyson equation and solves it using iterative
numerical methods, as described in Ref. \cite{mm}.  In Fig. \ref{Sigma_comp} we
show the solution for the dynamical fermion mass $\Sigma$ as a function of
$\alpha_*$.  A fit to the numerical solution in the walking region $0.89 \le
\alpha_* \le 1.0$ in Ref. \cite{mm} found agreement with the functional form
(\ref{sigsol}) with $c = 4.0$.  Our calculations for larger $\alpha_*$ show the
expected shift away from walking behavior.  This shift is evident in Fig.
\ref{Sigma_comp} for $\alpha_*$ larger than about 1.2.  Note that our solution
of the full Schwinger-Dyson equation does not make the approximation of setting
$\Sigma(q_E^2)=\Sigma$ in the fermion propagator denominator but instead
incorporates the full functional dependence of $\Sigma(q_E^2)$.  In real QCD,
precision fits to deep inelastic lepton scattering data, hadronic decays of the
$Z$, etc. probe the theory in momentum regions where $N_f = 4$ or $N_f=5$, and
yield, for the effective $N_f$-dependent scale $\Lambda^{(5)}_{QCD} \simeq 200$
MeV and $\Lambda^{(4)}_{QCD} \simeq 280$ MeV, with larger values for
$\Lambda^{(N_f)}_{QCD}$ with $N_f=3,2$. In actual QCD one thus has
$\Sigma/\Lambda^{(N_f)} \simeq O(1)$ for these low values of $N_f$.  These
contrast with the limiting walking behavior, in which $\Sigma \ll \Lambda$, as
indicated in eq. (\ref{sigsol}).  Our calculation of $\Sigma$, shown in
Fig. \ref{Sigma_comp}, shows that $\Sigma/\Lambda$ increases substantially, by
about a factor of 30, from a value of about 0.01 at $\alpha_* = 1.0$ to 0.32 at
$\alpha_*=2.5$, much closer to the value of O(1) for this ratio in QCD.

Another quantity of interest is the pseudoscalar decay constant $f_P$, the
$N_f$-flavor generalization of the pion decay constant.  For $N_f=2$ QCD this
is defined as $\langle 0 | J^j_\mu | \pi^k(q) \rangle = i f_\pi q_\mu
\delta^{jk}$ where $1 \le j,k \le 3$ are SU(2) isospin indices. Here, we use a
generalization of this definition, with the symbol $f_\pi$ replaced by $f_P$
and the SU($N_f$) isospin indices in the range $1 \le j,k \le N_f^2-1$.  In
QCD, one rough measure of the dynamical (constituent) quark mass is $\Sigma
\simeq M_N/N_c \simeq 313$ MeV, where $M_N$ is the nucleon mass.  An alternate
definition sets $\Sigma \simeq M_\rho/2$; this would yield a somewhat larger
value.  Here we use the estimate $\Sigma \simeq 330$ MeV.  Using the measured
value $f_\pi \simeq 92.4 \pm 0.3$ MeV \cite{pdg}, one thus has
\beq
\bigg (\frac{\Sigma}{f_\pi} \bigg )_{QCD} \simeq 3.6 \ . 
\label{sig_over_fpi_qcd}
\eeq

An approximate relation connecting $\Sigma$ and $f_P$ is \cite{psrel} (with $y
\equiv k_E^2$)
\beq
 f_P^2 \, = \, \frac{N_c}{4\pi^2} \int_0^\infty y \, dy \,
       \frac{\Sigma^2(y) \, - \,  \frac{y}{4} \,
\frac{d}{dy} \left[ \Sigma^2(y) \right] }
       { [ y \ +\  \Sigma^2(y) \ ]^2}  \ .
\label{psrel}
\eeq
The integration is rendered finite by the softness of the dynamical
mass $\Sigma(k_E^2)$, which behaves for $k \gg \Sigma$ as
\beq
\Sigma(k_E^2) \propto \Sigma \bigg ( \frac{\Sigma}{k_E} \bigg )^{2-\gamma}
\label{sigmak}
\eeq
where $\gamma$ is the anomalous dimension of the bilinear operator $\bar f f$,
having the value $\gamma \simeq 1$ in the walking regime and decreasing toward
zero at very large energy scales $\mu >> \Lambda$ (since $\gamma$ is
a power series in $\alpha$ and $\alpha \to 0$ in this limit due to the
asymptotic freedom of the theory).  Thus, the relation (\ref{psrel}) suggests
that for QCD 
\beq
f_\pi^2 \simeq \frac{N_c \Sigma^2}{4\pi^2}  \ . 
\label{fsigrel}
\eeq
For $N_c=3$, this is $\Sigma/f_\pi \simeq 2\pi/\sqrt{3} \simeq 3.6$, which
agrees, to within the theoretical uncertainties, with experiment.  In QCD,
with $\Lambda^{(2)} \simeq 400$ MeV, one has 
\beq
\frac{f_\pi}{\Lambda^{(2)}_{QCD}} \simeq 0.25 \ . 
\label{fpi_over_lam_qcd}
\eeq

\begin{figure}
  \begin{center}
    \includegraphics[height=6cm]{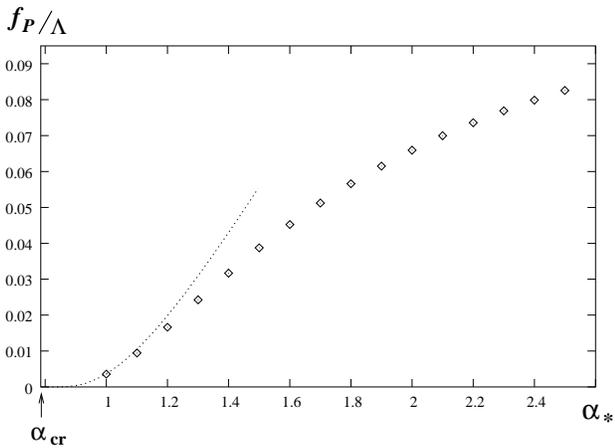}
  \end{center}
\caption{Values of $f_P$ calculated from eq. (\ref{psrel}) for several
  values of $\alpha_*$ (indicated by $\Diamond$).  For comparison, we show, as
  the dotted line, the analytic solution given by eq. (\ref{sigsol}) with
  $\Sigma$ replaced by $f_P$ and $c=1.5$ derived from a fit to the
  calculations for $0.89 \le \alpha_* \le 2.5$.  See text for further
 discussion.}
\label{fp_comp}
\end{figure}

In Fig. \ref{fp_comp} we show our results for $f_P$ calculated from
substituting our solution for $\Sigma(k^2)$ into eq. (\ref{psrel}).  In the
walking limit, $f_P$ has been shown to satisfy a relation similar to
eq. (\ref{sigsol}), i.e., it is exponentially smaller than the scale $\Lambda$.
We display, as the dotted curve, the fit from Ref. \cite{mm} for the walking
interval $0.89 \le \alpha_* \le 1.0$, given by eq. (\ref{sigsol}) with $c=1.5$.
Our results show the change from this walking type of behavior as $\alpha_*$
increases above 1.2; as $\alpha_*$ increases from 1.0 to 2.5, $f_P/\Lambda$
increases substantially, from about $ 3 \times 10^{-3}$ to about 0.08.  This is
similar to the factor by which we found that $\Sigma\Lambda$ increased as
$\alpha_*$ increased through this interval.

\begin{figure}
  \begin{center}
    \includegraphics[height=6cm]{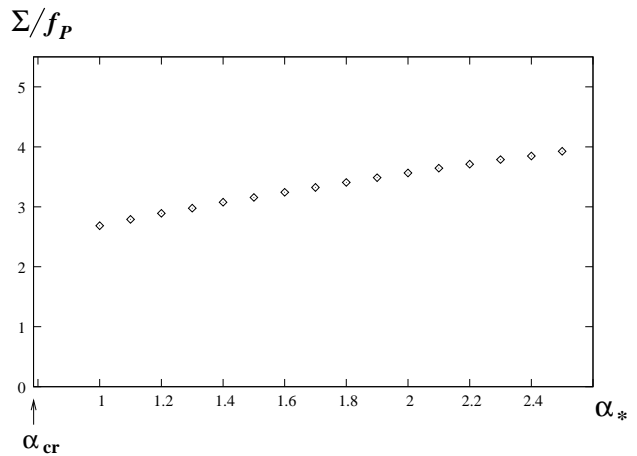}
  \end{center}
\caption{Plot of the ratio $\Sigma/f_P$ for $1 \le \alpha_* \le 2.5$.}
\label{Sigoverf}
\end{figure}

The strong increase in $\Sigma/\Lambda$ and $f_P/\Lambda$ as $\alpha_*$ ascends
from the value 0.89 near the walking limit to the value 2.5 deeper within the
confinement phase is easily understood as reflecting the removal of the extreme
exponential suppression evident in eq. (\ref{sigsol}) and its analogue for
$f_P$ for $\alpha_* - \alpha_{cr} \to 0^+$.  One does not expect such a
dramatic change in the ratio $\Sigma/f_P$ over this interval, and this
expectation is borne out by our calculations. In Fig. \ref{Sigoverf} we show
the ratio of $\Sigma/f_P$, which increases gradually from about 2.6 to 3.9.
The fact that we find a ratio comparable to the observed one in actual QCD,
given by eq. (\ref{sig_over_fpi_qcd}), can be understood as a consequence of
the property that much of the strong dependence on $N_f$ divides out in this
ratio.

\section{Calculation of Meson Masses via the Bethe-Salpeter Equation}

\subsection{General Discussion}

We denote the wavefunction for a hadron with a given flavor combination for the
generalized $\pi$, $\rho$, etc. as follows.  Define the flavor vector $f^a
\equiv (f^{a1},...,f^{aN_f})$.  Recall that in the confined phase the global
symmetry ${\rm SU}(N_f)_L \times {\rm SU}(N_f)_R \times U(1)_V$ is broken
spontaneously to ${\rm SU}(N_f)_V \times {\rm U}(1)_V$.  We drop the explicit
subscript $V$ on ${\rm SU}(N_f)_V$ henceforth.  With regard to ${\rm SU}(N_f)$,
a $f \bar f$ meson with a given $J^{PC}$ (where $J$ denotes the spin, and $P$
and $C$ are the parity and charge conjugate eigenvalues) is described via the
Clebsch-Gordan decomposition $N_f \times \bar N_f = 1 + Adj$, where 1 and $Adj$
denote the singlet and adjoint representations.

Let the generators of the group ${\rm SU}(N_f)$ have the standard normalization
${\rm Tr}(T_iT_j)=(1/2)\delta_{ij}$.  Then the hadrons transforming according
to the adjoint representation of ${\rm SU}(N_f)$ are comprised of (i) the set
of $N_f(N_f-1)$ states
\beq 
h_{\Gamma;ij} = \frac{1}{\sqrt{N_c}}\sum_{a=1}^{N_c}\bar f_a \, 
\Gamma \, T_{ij} \, f^a
\label{hij}
\eeq
where $T_{ij}$ is the $N_f \times N_f$ matrix with a 1 in the $i$'th column and
$j$'th row, with $1 \le i,j \le N_f$, $i \ne j$, and $\Gamma$ specifies the
type of particle (pseudoscalar, vector, axial-vector, scalar), and
(ii) the $N_f-1$ states corresponds to the generators of the Cartan subalgebra
of ${\rm SU}(N_f)$ given by the traceless $N_f \times N_f$ matrices
\beq 
T_{dk} = [2k(k+1)]^{-1/2}{\rm diag}(1,1,..,1,-k,0,..,0)
\label{tdk} 
\eeq 
where there are $k$ 1's and $1 \le k \le N_f-1$.  That is, $T_{d1}=(1/2){\rm
diag}(1,-1,...,0)$, $T_{d2}=(2\sqrt{3})^{-1}{\rm diag}(1,1,-2,0,...,0)$, etc.
Because of the ${\rm SU}(N_f)$ flavor symmetry, it does not matter which of
these $N_f^2-1$ hadrons with a given $\Gamma$ we use.  We shall refer to these
as the $N_f$-generalized $\rho$, $\omega$, etc.  In particular, the spectrum of
mesons includes a set of $N_f^2-1$ Nambu-Goldstone bosons (NGB's) with $L=S=0$
and $J^{PC}=0^{-+}$, transforming according to the adjoint representation of
${\rm SU}(N_f)$. The corresponding $0^{-+}$ singlet with respect to ${\rm
SU}(N_f)$, i.e., the generalized $\eta'$, is not a Nambu-Goldstone boson
because the corresponding U(1)$_A$ symmetry is anomalous.  Our analysis of
meson masses is for the lowest-lying $f \bar f$ states.  In future work one
could also consider radial excitations, Regge recurrences, pure gluonic states
(glueballs) and the general coupled situation in which glueballs and $f \bar f$
mesons of the same $J^{PC}$ mix.

In QCD, there are several (light-quark) $\bar q q$ mesons that are of interest
here. For the reader's convenience, we list these in Table \ref{mesons}.  A
notation for the various states in the case of general $N_f$ massless quarks is
$S_R$, $P_R$, $V_R$, and $A_R$, standing for ``scalar, pseudoscalar, vector,
and axial-vector'', where the subscript $R$ denotes the representation -
adjoint or singlet - under the SU($N_f$) flavor symmetry group.  The
experimental and theoretical situation concerning the $0^{++}$ isoscalar meson
$f_0$ has been the subject of much discussion over the years; indeed, this
state may involve mixing with $q q \bar q \bar q$ mesons \cite{schechter}.
Because of the complications in the analysis of this state, and the expected
complications in a realistic analysis of its $N_f$-generalization, the
SU($N_f$)-singlet $0^{++}$ meson, we do not attempt to treat this in our
current study.

\begin{table}
\caption{\footnotesize{Data on relevant $q \bar q$ mesons whose masses are
compared with Bethe-Salpeter calculations. $n {}^{2S+1} \, L_J$
is standard spectroscopic notation, where $n$ denotes radial quantum number.
The symbols adj. and sing. denote the adjoint and singlet representations of
the SU(2)$_V$ isospin flavor symmetry group.  Masses are given in MeV, from 
\cite{pdg}. The last column lists the mass divided by a typical hadronic scale,
$f_\pi$.}}
\begin{center}
\begin{tabular}{|c|c|c|c|c|c|c|} \hline
$n {}^{2S+1} \, L_J$ & $J^{PC}$ & $R_{SU(2)_V}$ & name &  $M$ & $M/f_\pi$ 
\\ \hline
$1 {}^3 \, S_1$  & $1^{--}$  & adj.  & $\rho$   & $775.8 \pm 0.5$ & 8.40 
\\ \hline
$1 {}^3 \, S_1$  & $1^{--}$  & sing. & $\omega$ & $782.6 \pm 0.1$ & 8.47
\\ \hline
$1 {}^1 \, P_1$  & $1^{+-}$  & adj.  & $b_1$    & $1229.5 \pm 3.2$ & 13.3
\\ \hline
$1 {}^1 \, P_1$  & $1^{+-}$  & sing.  & $h_1$    & $1170 \pm 20$   & $12.7 \pm
0.2$ \\ \hline
$1 {}^3 \, P_0$  & $0^{++}$  & adj.  & $a_0$    & $984.7 \pm 1.2$ & 10.7
\\ \hline
$1 {}^3 \, P_0$  & $0^{++}$  & sing. & $f_0$    & $\sim 600^{+600}_{-200}$ &
$6.5^{+6.5}_{-4.3}$ \\ \hline
$1 {}^3 \, P_1$  & $1^{++}$  & adj.  & $a_1$    & $1230 \pm 40$ & $13.3 \pm 
0.4$  \\ \hline
$1 {}^3 \, P_1$  & $1^{++}$  & sing. & $f_1$    & $1281.8 \pm 0.6$ & 13.9
\\ \hline
\end{tabular}
\end{center}
\label{mesons}
\end{table}

As will be seen below, in the Bethe-Salpeter equation that we use to calculate
the masses of the mesons, the flavor-dependent structure is simply
a prefactor.  Hence, the solutions of this equation have the property 
that, for a given radial quantum number and spectroscopic form 
${}^{2S+1} \, L_J$, the SU($N_f$) flavor-singlet and flavor-adjoint mesons 
are degenerate: 
\beq
M(n {}^{2S+1} \, L_J; {\rm flav. \ adjoint}) = 
M(n {}^{2S+1} \, L_J; {\rm flav. \ singlet})
\label{mrel}
\eeq
In view of this, we henceforth drop the subscript $R$ and simply write $V$
rather than $V_{flav. adj.}$ or $V_{flav. sing.}$, etc.  Note that this is
different from the prediction from SU($N_f$) flavor symmetry (with degenerate
quarks and electroweak interactions turned off) that the members of a given
representation of SU($N_f$) are degenerate.  Experimentally, except for the
pseudoscalar mesons, the light-quark isospin-adjoint and isospin-singlet $q
\bar q$ mesons are nearly degenerate. The physical $\omega$ meson is very
nearly a singlet under isospin SU(2), so a measure of this predicted degeneracy
for the ground state $1^{--}$ mesons is $(M_\omega - M_\rho)/[(1/2)((M_\omega +
M_\rho)] = 0.87 \times 10^{-2}$, quite small.  Similarly, $(M_{f_1} -
M_{a_1})/[(1/2)((M_{f_1} + M_{a_1})] = 0.04 \pm 0.03$ and $(M_{h_1} -
M_{b_1})/[(1/2)((M_{h_1} + M_{b_1})] =-0.05 \pm 0.02$.  So for these states the
prediction from our Bethe-Salpeter technique for the special case $N_f=2$
massless quarks is in agreement with the data for light-quark mesons in QCD.

The situation with the $0^{-+}$ mesons is quite different.  Since the SU($N_f$)
flavor-singlet mesons are not NGB's, owing to the anomalous nature of the
U(1)$_A$ symmetry, they are split by a large mass difference from the
flavor-adjoint NGB's.  In this case, as noted above, the semi-perturbative
Bethe-Salpeter analysis does not contain the relevant physics involving
instantons, and hence its prediction is far from reality.  For this reason we
do not consider the flavor-singlet $0^{-+}$ mesons here.  As regards the
flavor-adjoint $0^{-+}$ mesons, since we assume massless fermions and have
turned off electroweak interactions, the mass $M_P$ of the flavor-adjoint
pseudoscalar mesons is exactly zero in our calculations.  

The pion decay constant $f_\pi$ provides a convenient mass scale with which to
normalize the hadron masses.  For comparison with our results calculated in
the case of general larger $N_f$, we list in Table \ref{mesons} the masses of
the $q \bar q$ mesons divided by $f_\pi$.  For later use we also record the
ratio
\beq
\frac{M_{a_1}}{M_\rho} = 1.59 \pm 0.05 \ . 
\label{ma1mrho_qcd}
\eeq
This is slightly larger than the prediction $M_{a_1}/M_\rho = \sqrt{2} \simeq
1.414$ from a combination of vector meson dominance and spectral function sum
rules \cite{wein}.  Also,
\beq
\frac{M_{a_0}}{M_\rho} = 1.27 \ . 
\label{ma0mrho_qcd}
\eeq

An interesting result of the calculations of meson masses in the walking limit
in Ref. \cite{mm} was that the ratios of these masses to $f_P$ are rather
constant.  Specifically, it was found that in for $0.89 \le \alpha_* \le 1.0$, 
\beq
\frac{M_V}{f_P} \simeq 11 \ , 
\label{mvfp_ratio_wqcd}
\eeq
\beq
\frac{M_A}{f_P} \simeq 11.5 \ , 
\label{mafp_ratio_wqcd}
\eeq
and
\beq
\frac{M_S}{f_P} \simeq 4.1 \ , 
\label{msfp_ratio_wqcd}
\eeq
so that 
\beq
\frac{M_A}{M_V} = 1.04 
\label{mamvratio}
\eeq
and
\beq
\frac{M_S}{M_V} = 0.36 \ , 
\label{msmvratio}
\eeq
where the theoretical uncertainty is several per cent.  These ratios may be
compared with the values in regular QCD which, as far as the light-meson
spectrum is concerned, are close to the values that they would have in the
$N_f=2$ chiral limit (with the understanding that the pion masses would
actually vanish in this limit if electroweak interactions are turned off, as
assumed here).  For the purpose of this comparison, we do not try to use the
inferred chiral-limit value of $f_\pi$ \cite{fpi}, since to be consistent we
would have to do the same for the mesons themselves.  For the comparison
between the extreme walking limit (WL) and QCD, we have
\beq
\frac{(M_V/f_P)_{WL}}{(M_\rho/f_\pi)} \simeq 1.3
\label{mvratio_wq}
\eeq
\beq
\frac{(M_A/f_P)_{WL}}{(M_{a_1}/f_\pi)} \simeq 0.86 \ , 
\label{maratio_wq}
\eeq
and
\beq
\frac{(M_S/f_P)_{WL}}{(M_{a_0}/f_\pi)} \simeq 0.38 \ . 
\label{rsratio_wq}
\eeq
A major output of the present work is the elucidation of how, as $N_f$
decreases and $\alpha_*$ increases, the ratios of meson masses to $f_P$ begin 
to shift toward their QCD values.

\section{Calculations} 

Next, we describe the details of our solution of the Bethe-Salpeter equation
and the resulting masses of $f \bar f$ mesons.

\subsection{Bethe-Salpeter amplitudes}

We introduce the Bethe-Salpeter amplitudes $\chi$ for 
the scalar ($S$), pseudoscalar ($P$), vector ($V$), and 
axial-vector ($A$) bound states of quark and anti-quark as follows : 
\beqs
  \langle 0 \vert 
   T \psi_{\alpha f i}(x_+)\ \bar\psi_\beta^{f^\prime j}(x_-) 
  \ \vert S_a(q) \rangle  && \\
  = \sqrt{2} \, \delta_i^j (T_a)_f^{f^\prime} \, 
  e^{-iq \cdot X} \int \frac{d^4p}{(2 \pi)^4} 
e^{-ip \cdot r} &[\chi_{(S)}(p;q)]_{\alpha\beta}& , \nonumber 
\eeqs
\beqs
  \langle 0 \vert 
    \ T\  \psi_{\alpha f i}(x_+)\ \bar\psi_\beta^{f^\prime j}(x_-) 
  \ \vert P_a(q) \rangle && \\
  = \sqrt{2} \, \delta_i^j (T_a)_f^{f^\prime} \, 
  e^{-iq \cdot X} \int \frac{d^4p}{(2 \pi)^4} 
e^{-ip \cdot r} &[\chi_{(P)}(p;q)]_{\alpha\beta}& ,  \nonumber 
\eeqs
\beqs
  \langle 0 \vert 
    \ T\ \psi_{\alpha f i}(x_+)\ \bar\psi_\beta^{f^\prime j}(x_-) 
  \ \vert V_a(q,\epsilon) \rangle && \\
  = \sqrt{2} \, \delta_i^j (T_a)_f^{f^\prime} \, 
  e^{-iq \cdot X} \int \frac{d^4p}{(2 \pi)^4} 
e^{-ip \cdot r} &[\chi_{(V)}(p;q,\epsilon)]_{\alpha\beta}&,  \nonumber 
\eeqs
\beqs
  \langle 0 \vert 
    \ T\  \psi_{\alpha f i}(x_+)\ \bar\psi_\beta^{f^\prime j}(x_-) \ 
  \vert A_a(q,\epsilon) \rangle && \\
  = \sqrt{2} \, \delta_i^j (T_a)_f^{f^\prime} \, 
  e^{-iq \cdot X} \int \frac{d^4p}{(2 \pi)^4} 
e^{-ip \cdot r} &[\chi_{(A)}(p;q,\epsilon)]_{\alpha \beta}&, \nonumber 
\eeqs
where $x_\pm = X \pm r/2$, and ($\alpha$, $\beta$), ($f$, $f^\prime$), and
($i$, $j$) denote the spinor, flavor, and color indices, respectively.
$\lambda_a$ represents flavor structure of the bound states.  In the case of
flavor-adjoint bound states, $T_a$ is the generator of $\mbox{SU}(N_f)$,
while in the case of flavor singlet bound states, $(T_a)_{f}^{f'}$ is the
identity {\bf 1}.

We expand the BS amplitude $\chi$ in terms of the bispinor bases 
$\Gamma^i$ and the invariant amplitudes $\chi^{(i)}$ as follows :
\beq
\left[ \chi_{(S,P)}(p;q) \right]_{\alpha \beta} 
=\ \sum_{i=1}^{4} \left[ \Gamma^i_{(S,P)}(p;q) \right]_{\alpha \beta}
  \chi_{(S,P)}^{(i)} (p;q) ,
\label{eq:chi-SP}
\eeq
\beq
\left[ \chi_{(V,A)}(p;q,\epsilon) \right]_{\alpha \beta} 
=\ \sum_{i=1}^{8} \left[ \Gamma^i_{(V,A)}(p;q,\epsilon) \right]_{\alpha \beta}
  \chi_{(V,A)}^{(i)} (p;q) .
\label{eq:chi-VA}
\eeq
The bispinor bases can be determined from the spin, parity, and charge
conjugation properties of the bound states.  The explicit forms of
$\Gamma_{(S)}^i$, $\Gamma_{(P)}^i$, $\Gamma_{(V)}^i$, and $\Gamma_{(A)}^i$ are
summarized in Appendix~\ref{app:bispinor-bases}.

We take the rest frame of the bound state as a frame of reference:
\beq
  q^\mu = ( M_B , 0 , 0 , 0 ) ,
\eeq
where $M_B$ represents the bound state mass, i.e., $M_S, \ M_P, \ M_V$ and
$M_A$ for scalar, pseudoscalar, vector and axial-vector meson masses,
respectively.  After a Wick rotation, we parametrize $p^\mu$
by the real variables $u$ and $x$ as
\beq
  p \cdot q = i M_B u \ ,\  p^2 = - u^2 - x^2 .
\eeq
Consequently, the invariant amplitudes $\chi^{(i)}$
can be expressed as functions of the variables $u$ and $x$:
\beq
  \chi^{(i)}_{(S,P,V,A)} = \chi^{(i)}_{(S,P,V,A)}(u,x) .
\eeq
{}From the charge conjugation properties for the BS amplitude $\chi$ and the
bispinor bases defined in Appendix~\ref{app:bispinor-bases}, the invariant
amplitudes $\chi^{(i)}(u,x)$ are shown to satisfy the following relation:
\beq
  \chi^{(i)}_{(S,P,V,A)}(u,x) = \chi^{(i)}_{(S,P,V,A)}(-u,x)\ .
\label{eq:even_chi}
\eeq
%

\subsection{Homogeneous Bethe-Salpeter equation}

\begin{figure}
  \begin{center}
    \includegraphics[height=2.5cm]{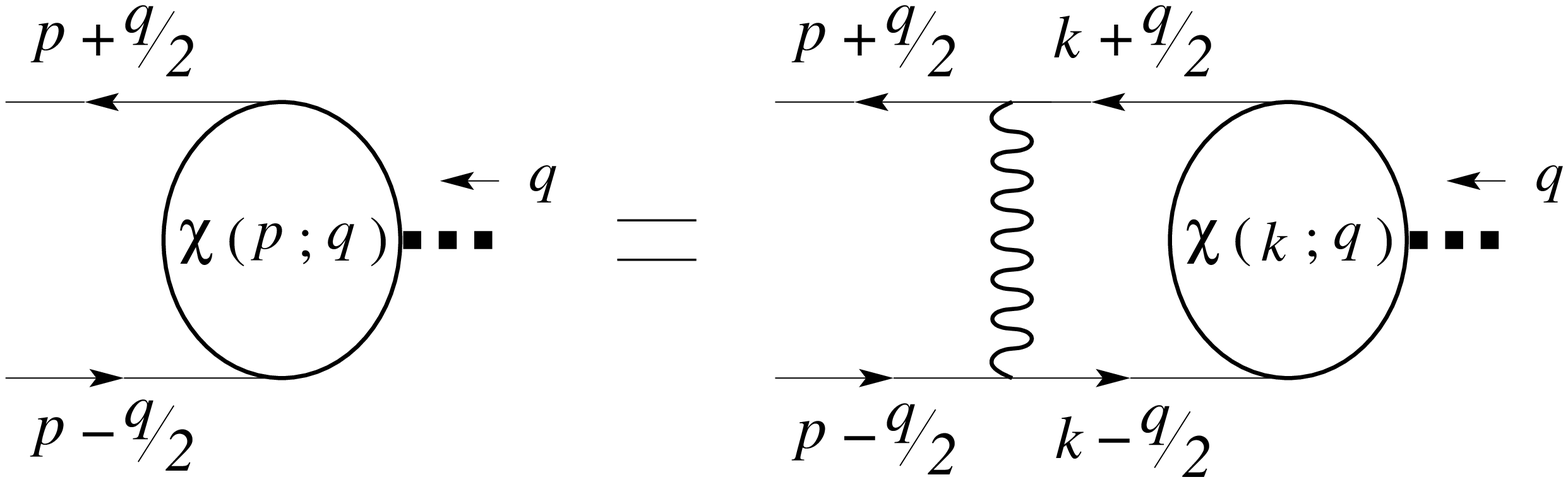}
  \end{center}
\caption{A graphical representation of the HBS equation 
in the (improved) ladder approximation}
\label{fig:HBSeq}
\end{figure}

The homogeneous Bethe-Salpeter (HBS) equation is the self-consistent equation
for the Bethe-Salpeter amplitude, and it is expressed as (see
Fig.~\ref{fig:HBSeq})

\beq
  T \chi = K \chi \ .
\label{eq:HBSeq}
\eeq
The kinetic part $T$ is given by 
\beq
  T(p;q) =  S_f^{-1}(p + q/2) \otimes S_f^{-1}(p - q/2) \ ,
\label{def T}
\eeq
where the BS kernel $K$ in the improved ladder approximation 
is expressed as
\beq
  K(p;k) \ =\    C_{2f} 
         \frac{4\pi \alpha(p,k)}{(p-k)^{2}}
         \left( g_{\mu\nu} - \frac{(p-k)_\mu (p-k)_\nu}{(p-k)^2}
         \right) \gamma^\mu \otimes \gamma^\nu .
\eeq
In the above expressions we used the tensor product notation
\beq
  (A \otimes B) \,\chi  =  A\, \chi\, B \ ,
\eeq
and the inner product notation 
\beq
   K \chi\ (p;q) = -i \int \frac{d^4 k}{(2\pi)^4}\  K(p,k)\  \chi(k;q) \ .
\eeq
It should be noted that the fermion propagators included in $T$ in
eq.~(\ref{def T}) have complex-valued arguments after the Wick rotation
\cite{sdcomplex}.  The arguments of the mass functions appearing in the two
legs of the Bethe-Salpeter amplitude are expressed as
\beq
  - ( p \pm q/2 )^2 = u^2 + x^2 - \left( \frac{M_B}{2} \right)^2 
                      \mp i u M_B .
\eeq
In general, it is difficult to obtain mass functions for complex arguments by
solving the Schwinger-Dyson equation in the complex plane, especially because
of the analytic structure of the running coupling in the complex momentum
plane.  However, in the case of large $N_f$ QCD, the analyticity of the
two-loop running coupling constant~\cite{Gardi} makes it possible to solve for
the mass function in the complex plane. This leads to the following
approximation, in accordance with eq.~(\ref{stepfun}) \cite{mm}:
\beqs
  {\rm Re}\left[\alpha(X e^{i2\theta})\right]
&=& \alpha_\ast\ \theta(\Lambda^2 - X),\label{eq:real}\\
  {\rm Im}\left[\alpha(X e^{i2\theta})\right]
&=& 0.
\eeqs
where no confusion should result from the use of the symbol $\theta$ on the
right-hand side of eq.~(\ref{eq:real}) as the step function. 
.

\subsection{Numerical results}

\begin{figure}
  \begin{center}
    \includegraphics[height=7cm]{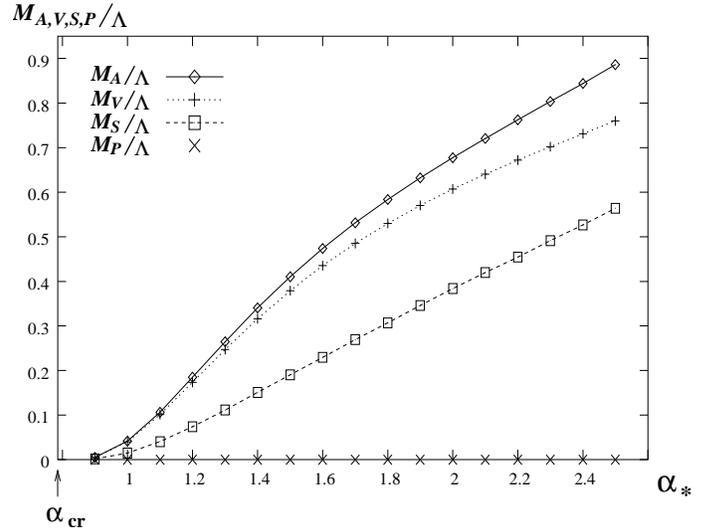}
  \end{center}
\caption{Values of meson masses divided by $\Lambda$ calculated from the
 Schwinger-Dyson and Bethe-Salpeter equations.}
\label{m_over_lam}
\end{figure}

\begin{figure}
  \begin{center}
    \includegraphics[height=7cm]{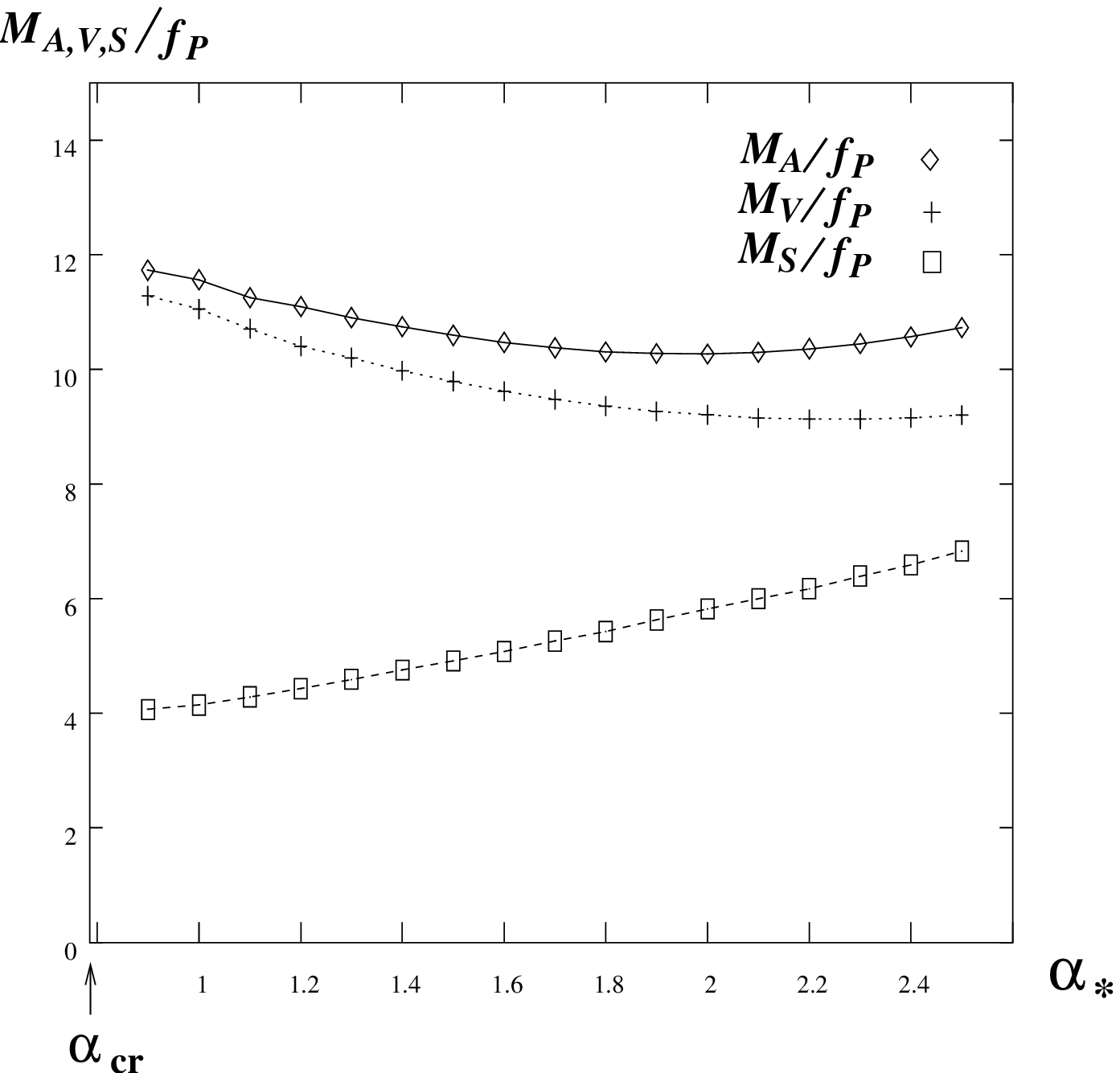}
  \end{center}
\caption{Values of meson masses divided by $F_P$ calculated from the
 Schwinger-Dyson and Bethe-Salpeter equations.}
\label{m_over_fp}
\end{figure}

\begin{figure}
  \begin{center}
    \includegraphics[height=7cm]{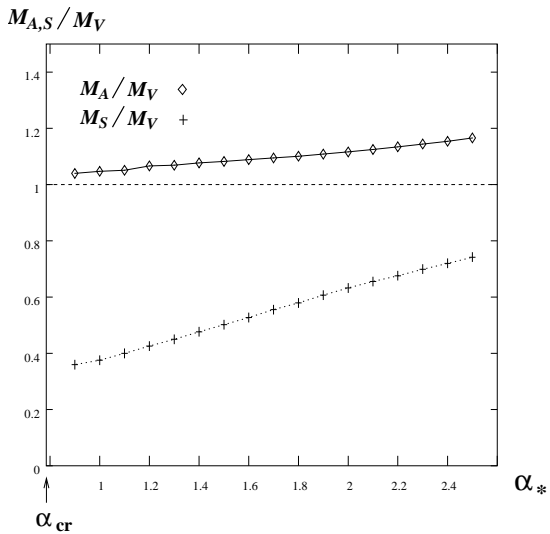}
  \end{center}
\caption{Ratios of meson masses calculated from the
 Schwinger-Dyson and Bethe-Salpeter equations.}
\label{mmratio}
\end{figure}

We next present the results of the numerical calculations for the masses of the
mesons.  We solve the homogeneous Bethe-Salpeter equation as an eigenvalue
problem, namely, the Bethe-Salpeter amplitude as an eigenfunction and the mass
of a bound state as an eigenvalue, denoted generically as $M_B$.  Because the
eigenvalue $M_B$ appears nonlinearly in the  equation, we use so-called
fictitious eigenvalue method \cite{HY} to obtain the value of $M_B$.  For
details of numerical method to solve HBS equation, see Ref.~\cite{mm}.  In
Fig.~\ref{m_over_lam}, we show the values of meson masses divided by $\Lambda$
calculated from the Schwinger-Dyson and Bethe-Salpeter equations in the range
$0.9 \le \alpha_* \le 2.5$.  In Fig.~\ref{m_over_fp} we plot the values of
$M_B/f_P$ in the range of $0.9 \le \alpha_* \le 2.5$.  In Fig.~\ref{mmratio},
we plot the meson mass ratios $M_A/M_V$ and $M_S/M_V$ in the range of $0.9 \le
\alpha_* \le 2.5$.


Our calculations yield a number of interesting results.  We summarize these
for the changes in these meson masses as $\alpha_*$ increases from 0.9 to 2.5
as follows.  

\begin{enumerate}

\item 

The ratios of the meson masses divided by $\Lambda$ increase dramatically, by
factors of order $10^2$, approaching values of order unity at $\alpha_* = 2.5$.
This amounts to the removal of the exponential suppression of these masses
which had described the walking limit at the boundary of the non-Abelian phase,
as one moves away from this limit into the interior of the confined phase. 

\item 

$M_S/f_P$ increases monotonically from about 4 to 7, thereby approaching to
within about 35 \% of the value 10.7 in QCD for $M_{a_0}/f_\pi$.

\item 

$M_V/f_P$ decreases from about 11 to 9, rather close to the value 8.5 for
$M_\rho/f_\pi$ and $M_\omega/f_\pi$ in QCD.  As is evident from
Fig. \ref{m_over_fp}, this ratio $M_V/f_P$ is roughly constant in the upper end
of the interval of $\alpha_*$ values that we study.

\item 

$M_A/f_P$ behaves non-monotonically, first decreasing from roughly 11.5 to 10,
but then increasing to about 11, within about 20 \% of the average of the
values in QCD for the isospin-triplet and isospin-singlet axial-vector mesons,
13 for $M_{a_1}/f_\pi$ and 14 for $M_{f_1}/f_\pi$.

\item 

Thus, the ratios $M_A/M_V$ and $M_S/M_V$, which were found in Ref. \cite{mm} to
have the respective values 1.04 and 0.36 in the walking limit, both increase in
the interval of $\alpha_*$ that we study, reaching about 1.17 and 0.74,
respectively, at $\alpha_*=2.5$.  For comparison, these ratios are
approximately 1.6 and 1.3 in QCD (cf. eqs.  (\ref{ma1mrho_qcd}) and
(\ref{ma0mrho_qcd})).  Although the value of the ratio $M_S/M_V$ at
$\alpha_*=2.5$ is farther from its QCD value than is the case with $M_A/M_V$,
it is increasing somewhat more rapidly as a function of $\alpha_*$, consistent
with eventually approaching the QCD value.

\end{enumerate}

\section{Conclusions} 

In this paper we have considered a vectorial SU($N_c$) gauge theory with $N_f$
massless fermions transforming according to the fundamental representation and
have studied the shift in behavior from walking that occurs in the region near
the boundary between the confinement phase and the non-Abelian Coulomb phase to
the QCD-like behavior with a non-walking coupling.  Specifically, we have used
the Schwinger-Dyson and Bethe-Salpeter equations to calculate the dynamically
induced fermion mass $\Sigma$, the spontaneous chiral symmetry breaking
parameter $f_P$, and the masses of the lowest-lying $q \bar q$ vector,
axial-vector, and flavor-adjoint scalar mesons.  We have investigated how these
change as one decreases $N_f$ below $N_{f,cr}$, or equivalently, increases
$\alpha_*$ above $\alpha_{cr}$, to move away from the above-mentioned boundary
into the interior of the confinement phase.  Our results show the crossover
between walking and non-walking behavior in a gauge theory.

There are a number of interesting topics for future research using the methods
of this paper.  It would be useful to construct a kernel for the Bethe-Salpeter
equation that could include more of the relevant physics, including instantons
effects.  Work is underway on this.  It would also be worthwhile to calculate
the masses of radially excited mesons and mesons with internal orbital angular
momenta $L \ge 2$, as well as glueballs and the mixing between glueballs and
$\bar q q$ mesons.  We anticipate that the results of these calculations would
exhibit the same general properties that we have found with the ground-state
$\bar q q $ mesons, but it would be interesting to confirm this expectation
explicitly. Another project would be to connect our study of the region in
$N_f$ where there is a crossover from walking to nonwalking behavior, to the
region around $N_f=2$. However, when one moves this far away from the walking
regime, one loses a simplifying features of our calculation, namely the fact
that we do not have to use an infrared cutoff on $\alpha$.  Given that lattice
gauge theory methods provide an {\it ab initio} framework for calculating
hadron masses, we hope that the lattice community will extend early efforts
such as those of Ref. \cite{mawhinney} and carry out a definitive study of
hadron masses in QCD with an arbitrary number of flavors.  It would be of
considerable interest to compare the results of the lattice calculations with
those obtained from solutions of Schwinger-Dyson and Bethe-Salpeter equations.

\acknowledgements

This research was partially supported by the grant NSF-PHY-00-98527.
M.K. thanks Profs. M. Harada and K. Yamawaki for the collaboration on the
related Ref. \cite{mm}, and R.S. thanks Dr. Neil Christensen for useful
comments.

\bigskip
\bigskip

\appendix

\section{Bispinor bases for scalar, pseudoscalar, vector, and 
axial-vector bound states}
\label{app:bispinor-bases}

In this appendix we show the explicit forms of the bispinor bases for the
scalar, pseudoscalar, vector, and axial-vector bound states.  Here we use the
notation $\hat{q}_\mu = q_\mu / M_B $ with $M_B$ being the mass of the bound
states, and $[a,b,c] \equiv a[b,c] + b[c,a] + c[a,b]$.

Bispinor base for the scalar bound state ($J^{PC} = 0^{++}$) 
is given by
\beqs
  \Gamma_{(S)}^1 &=&  {\bf 1} ,\ \ 
  \Gamma_{(S)}^2 =  \fsl{p} ,\ \ 
  \Gamma_{(S)}^3 =  \fsl{\hat{q}} (p \cdot \hat{q}) , \nonumber\\ 
  \Gamma_{(S)}^4 &=&  \frac{1}{2}\ [\fsl{p},\fsl{\hat{q}}] ,
\eeqs
and that for the pseudoscalar bound state ($J^{PC} = 0^{-+}$) 
is given by 
\beqs
  \Gamma_{(P)}^1 &=& \gamma_5 ,\ \ 
  \Gamma_{(P)}^2 = \fsl{p} \ (p \cdot \hat{q})\ \gamma_5 ,\ \ 
  \Gamma_{(P)}^3 = \fsl{\hat{q}} \ \gamma_5 , \nonumber\\
  \Gamma_{(P)}^4 &=& \frac{1}{2}\ [\fsl{p},\fsl{\hat{q}}]\ 
                   \gamma_5 \ .
\eeqs
Furthermore, for the vector bound state ($J^{PC} = 1^{--}$) we use
\beqs
&&  \Gamma_{(V)}^1 = \fsl{\epsilon} ,\ \ 
     \Gamma_{(V)}^2 = \frac{1}{2} [\fsl{\epsilon},\fsl{p}] 
                           (p \cdot \hat q) ,\ \ 
     \Gamma_{(V)}^3 = \frac{1}{2} [\fsl{\epsilon},\fsl{\hat q}] ,\nonumber\\
&&     \Gamma_{(V)}^4 = \frac{1}{3!}[\fsl{\epsilon},\fsl{p},\fsl{\hat
     q}] , 
     \Gamma_{(V)}^5 = (\epsilon \cdot p) ,\ \ 
     \Gamma_{(V)}^6 = \fsl{p} (\epsilon \cdot p) ,\nonumber\\
&&   \Gamma_{(V)}^7 = \fsl{\hat q}(p \cdot \hat q) (\epsilon \cdot p) ,\ \ 
     \Gamma_{(V)}^8 = \frac{1}{2} [\fsl{p},\fsl{\hat q}](\epsilon \cdot p) ,
\label{V bases}
\eeqs
and for the axial-vector bound state  
($J^{PC} = 1^{++}$) 
\beqs
&&   \Gamma_{(A)}^1 = \fsl{\epsilon}\ \gamma_5 ,\ \ \ 
     \Gamma_{(A)}^2 = \frac{1}{2} [\fsl{\epsilon},\fsl{p}] 
                           \gamma_5  ,\ \ \ 
     \Gamma_{(A)}^3 = \frac{1}{2} [\fsl{\epsilon},\fsl{\hat q}]\ 
        (p \cdot \hat q) \ \gamma_5  ,\nonumber\\
&&   \Gamma_{(A)}^4 = \frac{1}{3!}[\fsl{\epsilon},\fsl{p},\fsl{\hat q}]
     \ \gamma_5 ,\ \ \ 
     \Gamma_{(A)}^5 = (\epsilon \cdot p)\ (p \cdot \hat q)\ \gamma_5 ,\nonumber\\
&&   \Gamma_{(A)}^6 = \fsl{p} (\epsilon \cdot p)\ \gamma_5  ,\ \ 
     \Gamma_{(A)}^7 = \fsl{\hat q}\ (\epsilon \cdot p)\ (p \cdot \hat q)
                \ \gamma_5  ,\nonumber\\
&&   \Gamma_{(A)}^8 = \frac{1}{2} [\fsl{p},\fsl{\hat q}](\epsilon \cdot p)\
     (p \cdot \hat q)\ \gamma_5  .
\eeqs


\begin{thebibliography}{99}
\bibitem{integer}
%
Here and below, when we mention non-integral values of $N_f$, it is implicitly
understood that physical values of $N_F$ are, of course, positive integers, and
the non-integral values are defined via an analytic continuation away from
these physical values.
\bibitem{casimir} The Casimir invariant $C_2(R)$ of the representation $R$ is
defined by ${\cal D}_R(T_a)^i_j{\cal D}_R(T_a)^j_k = C_2(R)\delta^i_k$, where
$\{a,b\}$ and $\{i,j,k\}$ denote group and representation indices and sums over
repeated indices are understood.
\bibitem{fpi} 
It has been estimated \cite{gl85} that $f_\pi/(f_\pi)_{ch.lim.} \simeq 1.06$,
where $(f_\pi)_{ch.lim.}$ denotes the value of $f_\pi$ in the chiral 
limit $m_u=m_d=0$.  With $f_\pi=92.4$, this gives $(f_\pi)_{ch.lim.} \simeq 
87$ MeV.
\end{thebibliography}
\end{document}